\documentclass[aps,prd,superscriptaddress,twoside,twocolumn,nofootinbib,%
10pt,showpacs,floatfix]{revtex4-1}

\usepackage{amsmath,amssymb}
\usepackage{graphicx,bm}
\usepackage{slashed}
\usepackage{dcolumn}

\allowdisplaybreaks

\begin{document}


\title{Testing the tetraquark structure for the $X$ resonances in low-lying region}


\author{Hungchong Kim}%
\email{hungchong@kookmin.ac.kr}
\affiliation{Department of General Education, Kookmin University, Seoul 136-702, Korea}

\author{K. S. Kim}
\affiliation{School of Liberal Arts and Science, Korea Aerospace University, Goyang 412-791, Korea}

\author{Myung-Ki Cheoun}%
\affiliation{Department of Physics, Soongsil University, Seoul 156-743, Korea}

\author{Daisuke Jido}%
\affiliation{Department of Physics, Tokyo Metropolitan University, Hachioji, Tokyo 192-0397, Japan}

\author{Makoto Oka}%
\affiliation{Department of Physics, Tokyo Institute of Technology, Meguro 152-8551, Japan }
\affiliation{Advanced Science Research Center, Japan Atomic Energy Agency, Tokai, Ibaraki, 319-1195 Japan}

\date{\today}


\begin{abstract}

Assuming four-quark structure for the $X$ resonances in low-lying region, we calculate their masses using
the color-spin interaction.
In specific, the hyperfine masses of the color-spin interaction are calculated for the possible states
in spin-0, spin-1, spin-2 channels.
The two states in spin-0 channel as well as the two states in spin-1 channel are diagonalized
in order to generate the physical hyperfine masses.
By matching the difference in hyperfine masses with the splitting in corresponding hadron masses and using the $X(3872)$
mass as an input, we estimate the masses corresponding to the states $J^{PC}=0^{++}, 1^{+-},2^{++}$.
We find that the masses of two states in $1^{+-}$ are close to those of $X(3823)$, $X(3900)$, and the mass of
the $2^{++}$ state is close to that of $X(3940)$. For them, the discrepancies are about $\sim 10$ MeV.
This may suggest that the quantum numbers of the controversial states are $X(3823)=1^{+-}, X(3900)=1^{+-}, X(3940)=2^{++}$.
In this work, we use the same inputs parameters, the constituent quark masses and the strength of the color-spin
interaction, that have been adopted in the previous
work on the $D$ or $B$-meson excited states. There, it was shown that the four-quark structure can be manifested in their excited states.
Thus, our results in this work provide a consistent treatment on open- and hidden-charm mesons as far as the four-quark
model is concerned.

\end{abstract}


\maketitle

\section{Introduction}

The exotic resonances, commonly refereed to $XYZ$ particles, are interesting subjects in hadron physics in recent years.
This interest was triggered by the measurement of $X(3872)$ back in 2003 by the Belle collaboration~\cite{Belle03} whose existence was
further confirmed in the later
experiments~\cite{Aubert:2004zr, Choi:2011fc, Aaij:2013zoa}.

The theoretical issues on $X(3872)$ are mainly on its structure. The traditional picture would be the quarkonium where
it is viewed as heavy quark-antiquark bound state,
${\bar c}c$~\cite{Eichten:2004uh}. The other picture for $X(3872)$ is a hadronic molecule
like $DD^*$~\cite{Swanson:2004pp,Tornqvist:2004qy}, where the two mesons are bounded weakly by hadronic interactions.
There is also a hybrid model like the type ${\bar c}c g$~\cite{Li:2004sta} containing a gluon.
This type of the hybrid model
is also applied~\cite{Close:2005iz} to the other resonance with higher mass, $Y(4260)$, whose existence is experimentally reported
in Ref.~\cite{BABAR05b}.  Another exciting scenario for $X(3872)$ would be a
long-sought tetraquark. Along this line, the tetraquark possibility was investigated under a traditional picture based on a diquark-antidiquark
form~\cite{Maiani:2004vq,Maiani:2014aja}
as well as under a more extensive treatment~\cite{Zhao:2014qva,Cui:2006mp}.
As far as we know, there is no consensus on the structure among all these possibilities.
The recent review given in Ref.~\cite{Hosaka:2016pey} might be useful for current status of $XYZ$ spectroscopy and possible future direction.

Additional resonances reported from later experiments, which have similar masses with $X(3872)$, escalate the interest
further because of the expectation that they might share a similar structure.
These include $X(3823)$~\cite{Bhardwaj:2013rmw},
$X(3900)^\pm$~\cite{Ablikim:2013mio,Liu:2013dau}, $X(3900)^0$~\cite{Xiao:2013iha},
$X(3940)$~\cite{Abe:2007jna}. These resonances, including $X(3872)$, constitute low-lying resonances in this $X$ spectrum.
Of course, there are other resonances with higher masses, $Y(4260)$~\cite{BABAR05b}, $Z(4430)$~\cite{Belle08b} {\it etc},
which provide other exciting areas that should be investigated in future.

One practical problem, in dealing with these resonances especially in low-lying region, is that their quantum numbers are scarcely known.
According to the Particle Data Group (PDG)~\cite{PDG14}, the only resonance whose quantum numbers are completely determined is $X(3872)$.
The quantum numbers of others in this low-lying region are quite unknown as one can see from Table~\ref{X resonances}.
This precisely makes it difficult to construct a realistic model for them.
In particular, for $X(3823)^{0}$, most of the quantum numbers are unknown except its charge conjugation being odd.  One interpretation for this state is
an accompanying member of $\psi(3770)$ when $\psi(3770)$ is viewed as a $^3D_1$ of ${\bar c}c$. Namely, $X(3823)^{0}$ could be a $^3D_2$ state but, without confirming
its quantum numbers, the issue is not settled yet. The exotic possibility is still open for this resonance.
Therefore, at this stage, the available data on these resonances which can be used to test a model construction are their measured masses and decay modes.


\begin{table}
\centering
\begin{tabular}{c|c|c|c}  \hline\hline
 Meson  & ~~$I^G (J^{PC})$~~ & ~~Mass (MeV)~~ & ~~~$\Gamma$ (MeV) \\
 \hline
 $X(3823)^{0}$ & $??(?^{?-})$ & 3823.1 & $ < 24$   \\
 $X(3872)^{0}$ & $0^+(1^{++})$ & 3871.7 & $ < 1.2$  \\
 $X(3900)^{\pm}$ & $?(1^+)$ & 3888.7 & $ 35 $   \\
 $X(3900)^{0}$ & $?(?^{?})$ & 3904 & -  \\
 $X(3940)^0$ & $?^?(?^{??})$ & 3942 & 37  \\
\hline\hline
\end{tabular}
\caption{The $X$ resonances collected from PDG~\cite{PDG14} in low-lying region. The charge, denoted by a
superscript in the name of each resonance, is assigned based on its decaying channel. Note,
the measured masses in the third column are slightly different from the numbers appearing in the resonance names in the first column.
The measured masses in the third column will be used in our analysis.}
\label{X resonances}
\end{table}

In this work, we investigate the tetraquark possibility for the $X$ resonances in low-lying region.
Especially we look for such a possibility from the electrically neutral members, $X(3823)^0$, $X(3872)^0$, $X(3900)^0$, $X(3940)^0$
by assuming that they belong to the same isospin multiplet. Since $X(3872)$ is isoscalar ($I=0$),
this assumption is equivalent to the statement that the other resonances, $X(3823)^0$, $X(3900)^0$, $X(3940)^0$, are also isocalar.
Of course, according to the $SU(3)_f$ symmetry, one can expect their charged partners (isovector) in the same $SU(3)_f$ multiplet
whose existence can support our isoscalar framework more.
Currently there is only one candidate for them in the similar mass region, $X(3900)^\pm$, which is not enough
to support our isoscalar tetraquark fully. As more or more resonances are updating
in this scarcely explored region, one can expect that more isovector resonances might appear in future.
One possible explanation for the lack of the isovector resonances in the current PDG might due to their large width, which makes
them difficult to be measured.
In particular, the isovector resonances can decay to ${\bar c}c +\pi$ which is however hindered by the isospin conservation for the isoscalar $X$ resonances.

We introduce the four-quark states consisting of diquark-antidiquark, $cq{\bar c}{\bar q}, (q=u,d)$ and test their
justification by calculating the masses using the color-spin interaction.  The construction of diquark-antidiquark and its application in light-quark sectors can be
found in Ref.~\cite{Jaffe77a,Jaffe77b,Jaffe04}.
This type of four-quark model for the $X$ resonances was also studied by Maiani {\it et al.}~\cite{Maiani:2004vq, Maiani:2014aja}.
They introduced this type of
wave function and calculated the masses using the color-spin interaction. In their approach, the constituent quark masses as well
as the strength of the color-spin interaction are all fitted to the low-lying baryon and meson spectra.
What we want to do in this work is to upgrade this approach by implementing a few ingredients.

Firstly, we notice that the color-spin interaction alone
is only useful to calculate the mass splittings through the hyperfine mass splittings.
The color-spin interaction does not represent the full interaction between two quarks and, therefore,
may not be enough to calculate the hadron mass itself. In specific, there are additional potentials such as
the color-electric term and the constant shift
as will be discussed below.  What makes the color-spin interaction special is that, in the mass splittings, the color-spin
interaction survives while the additional potentials cancel.
Secondly, as we will discuss below, there is a physical issue like mixing in the possible spin states.
If one constructs spin states from $cq{\bar c}{\bar q}, (q=u,d)$, there are two states in spin-0 and three states in spin-1.
Some of them must mix in generating the mass eigenstates, which constitutes another ingredient in this development.

Another thing, which is more important to us, is that this type of approach is closely related to the earlier work~\cite{Kim:2014ywa}.
Namely, in Ref.~\cite{Kim:2014ywa} where
two authors (H.Kim and M.K.Cheoun) in the present paper are involved, we suggested that most of the $D$ or $B$-meson excited states
currently listed in PDG, especially
their mass spectrum, can be understood if they are viewed as tetraquarks with the diquark-antidiquark form, $cq{\bar q}{\bar q}, (q=u,d,s)$.
Using the strength of the color-spin interaction fixed from
the mass difference of $D^{*0}_2 (2463)- D^{*0}_0(2318)$, we were able to reproduce the masses of other resonances in the excited states of $D$ and $B$ mesons quite
successfully. Also our four-quark model provides interesting phenomenology related to decays of spin-1 mesons, which seems to fit nicely with
experimental observation.  Based on its phenomenological success, we made some predictions for the $D$ and $B$ mesons to be found in future.
This approach in Ref.\cite{Kim:2014ywa} can be straightforwardly applied to the hidden-charm case here.
Through this application, one can test the parameters fixed in Ref.~\cite{Kim:2014ywa} which can help to construct a
consistent four-quark picture that might be relevant for both, open- and hidden-charm mesons.

The paper is organized as follows.
In Sec.~\ref{sec:wf}, we present four-quark wave functions that could be relevant for
the $X$ resonances. The color-spin interaction and its application to the
hadron spectroscopy is discussed in Sec.~\ref{sec:color-spin}.
In Sec.~\ref{sec:results}, we present our
calculations of the hyperfine masses and discuss their implication in the $X$ spectroscopy.
We summarize in Sec.~\ref{sec:summary}.

\section{Wave functions for the $X$ resonances}
\label{sec:wf}

In this section, we introduce the four-quark wave functions which could be relevant for the $X$ resonances.
These wave functions, in our best recollection, was first used
by Maiani {\it et al.}~\cite{Maiani:2004vq, Maiani:2014aja}.  Here we follow their
argument closely but make a few statements which need to be implemented in this update.

The $X$ resonances are hidden-charm mesons whose decay modes mostly
entail $J/\psi$, $D {\bar D}$, $D {\bar D}^*$, {\it etc}, in their final states.
Based on this observation, one can assume that their wave functions in the four-quark picture take
the diquark-antidiquark form, $cq{\bar c}{\bar q}, (q=u,d)$. One can show that
the diquark (antidiquark) forms a bound state when it is in ${\bar{\bm{3}}}_c$ (${\bm 3}_c$) and in the spin state $J=0$~\cite{Jaffe77a,Jaffe77b,Jaffe04}.
These two colorful objects, a diquark and an antidiquark, can be combined to make a four-quark state.
Then following Refs.~\cite{Maiani:2004vq, Maiani:2014aja}, we take the approximate spin independence of the heavy quark interaction
to assume that
the $J=1$ diquark (as well as the $J=1$ antidiquark) also forms a
bound state.

For the other color configuration for the diquark $cq$, namely $\bm{6}_c$, we do not consider this possibility
firstly because this is less compact than
the ${\bar{\bm{3}}}_c$ $cq$ in the spin-0 channel.
Secondly, including the $\bm{6}_c$ scenario through the mixing with the ${\bar{\bm{3}}}_c$ doubles
the possible states, which do not seem to fit the current $X$ phenomenology.
More importantly, the similar tetraquark framework applied to the excited states of $D,D_s,B,B_s$ mesons in
Ref.~\cite{Kim:2014ywa}, where the ${\bar{\bm{3}}}_c$ diquark is adopted only, provides nice phenomenological consequences, such as mass splittings.
Moreover, this approach predicts two spin-1 resonances which fits very
 well with $D^\pm_{s1} (2460)$, $D^\pm_{s1} (2535)$.
It is our interest to test whether the similar framework works for the other hadrons like the $X$ resonances.

Instead of this type of diquark-antidiquark picture, one can also consider a four-quark structure
of the form $q\bar{q}-c\bar{c}$ because of the expectation that the color-spin interaction can be more
attractive for $q\bar{q}$ than for the ${\bar{\bm{3}}}_c$ $cq$.
Indeed, the color-spin interaction is most attractive when $q\bar{q}$ is
in a state with color singlet and spin zero, i.e., (${\bf 1}_c$, $J=0$). With this $q\bar{q}$,  $q\bar{q}$ and $c\bar{c}$ are both {\it colorless}.
This leads to a hadronic molecule of meson-meson bound state, which tents to be separate with a long-range interaction.

Now with the other possible color-spin configurations for $q\bar{q}$, (${\bf 8}_c$, $J=0$), (${\bf 8}_c$, $J=1$), one can still
construct tetraquarks of the form, $q\bar{q}-c\bar{c}$, because now $q\bar{q}$ and $c\bar{c}$ are both {\it colorful}.
Then the question is whether the $q\bar{q}$ with (${\bf 8}_c$, $J=0$), (${\bf 8}_c$, $J=1$) configurations are more compact than the ${\bar{\bm{3}}}_c$ $cq$.
One can check this explicitly by calculating the expectation value of the color-spin interaction with respect to
the $q\bar{q}$ and the ${\bar{\bm{3}}}_c$ $cq$ states respectively. Using the constituent quark masses adopted in our work,
we find that the ${\bar{\bm{3}}}_c$ $cq$ with $J=0$ is most attractive among them.
Thus, the diquark $cq$ with the color ${\bar{\bm{3}}}_c$ and spin $J=0$ is appropriate for the tetraquark study.

For the $J=1$ diquark $cq$ with ${\bar{\bm{3}}}_c$, its color-spin interaction is stronger than the $q\bar{q}$
with (${\bf 8}_c$, $J=0$) but weaker than the $q\bar{q}$ with (${\bf 8}_c$, $J=1$).  In this sense, it is not
clear whether the $J=1$ diquark is appropriate for the tetraquark study.
However, at the same time, since $c\bar{c}$ is less compact than $q\bar{q}$,
it is also not clear whether the composite system, $q\bar{q}-c\bar{c}$,
is easier to form than the $cq-{\bar c}{\bar q}$ when the $J=1$ diquark is involved.

As we stated above, we take the approximate spin independence of the heavy quark limit to assume that the $J=1$ diquark
forms a bound state. Note, however, that the calculation based on the diquark-antidiquark basis, $cq{\bar c}{\bar q}$,
does not rule out the $q\bar{q}-c\bar{c}$ configuration entirely. Since we
calculate the color-spin interactions among all the quark pairs [see Eq.(\ref{color spin}) below], the
$q\bar{q}$ components with the color-octet as well as the color-singlet are also included in our framework.

In our work, we assume that the $X$ resonances, $X(3823)^0$, $X(3872)^0$, $X(3900)^0$, $X(3940)^0$, are isoscalar.
Then the flavor structure would be like $[cu{\bar c}{\bar u}+cd{\bar c}{\bar d}]/\sqrt{2}$. Technically,
this is equivalent to considering the configuration of the form, $cu{\bar c}{\bar u}$, within our framework
facilitating the color-spin interaction.

For illustration purpose, we label the four quarks
in the state, $cu{\bar c}{\bar u}$, with the numerical indices {1234}.
So [12] refers the two quarks in the diquark $cu$ and [34] refers to the two quarks in the antidiquark ${\bar c}{\bar u}$.
Using this labeling,
we can explain our formalism more clearly on the one hand, and on the other hand,
the general expressions involving the indices can be easily extended to other situations in future.

Assuming that all the quarks are in an $S$-wave state, the possible spins for the four-quark states are $J=0,1,2$.
If we denote these spin states
in terms of the total spin $J$, the diquark spin $J_{12}$, the antidiquark spin $J_{34}$, {\it i.e.}, $|J,J_{12},J_{34}\rangle$,
then we come up with the possible spin configurations for each spin state as follow,
\begin{eqnarray}
&J=0&:  \quad |0 0 0\rangle , \quad |0 1 1\rangle ,
\nonumber \\
&J=1&:  \quad |1 1 1\rangle , \quad |1 1 0\rangle , \quad |1 0 1\rangle\ ,
\nonumber \\
&J=2&:   \quad |2 1 1\rangle ,
\end{eqnarray}
where all the states have the positive parity by their construction.  From these spin configurations,
one can investigate how they behave under charge conjugation~\cite{Maiani:2004vq, Maiani:2014aja}.
The states in $J=0$ and $J=2$ are even under charge conjugation $C$. $|1 1 1\rangle$ is odd under $C$. And
$|1 1 0\rangle$ and $|1 0 1\rangle$ transform among each other under $C$.
Using these properties, one can classify the spin configurations in terms of the states based on the quantum numbers $J^{PC}$,
which then can be identified with physical states easily.

For $J=2$, there is only one spin configuration which can be denoted as
\begin{eqnarray}
|2^{++}\rangle = |2 1 1\rangle \ .
\label{spin 2}
\end{eqnarray}
For $J=0$, we have two spin configurations and these have the same quantum numbers $J^{PC}=0^{++}$.
To distinguish the two states in $|J^{PC}\rangle$ notation, we further label them, using $a$ and $b$, as
\begin{eqnarray}
|0^{++}_a\rangle = |0 0 0\rangle , \quad  |0^{++}_b\rangle= |0 1 1\rangle\ .
\label{spin 0}
\end{eqnarray}
What we want to point out in this work is that these two are not the mass eigenstates.
Indeed, as we will show in Sec.~\ref{sec:results}, the mixing element of the color-spin interaction, $\langle 0^{++}_a | V |0^{++}_b\rangle$,
is not zero. Thus, the mass eigenstates, which can be recognized as the physical states,
should be linear combinations of the two in Eq.~(\ref{spin 0}).

For $J=1$, we have three spin configurations.  We can assign one of them or a combination of them to $X(3872)$.
$X(3872)$ has well-determined quantum numbers, {\it i.e.,} $J^{PC}=1^{++}$. Only possible way to generate this
state with $1^{++}$ from the three configurations is the following symmetric combination~\cite{Maiani:2004vq, Maiani:2014aja},
\begin{equation}
|1^{++} \rangle = \frac{1}{\sqrt{2}} [ |1 1 0\rangle + |1 0 1\rangle ]\ .
\label{1++}
\end{equation}
The antisymmetric combination of $|1 1 0\rangle$ and $|1 0 1\rangle$ generates the state belonging to $J^{PC}=1^{+-}$ and
the other spin-1 state, $|1 1 1\rangle$, also belongs to $J^{PC}=1^{+-}$.
Thus, we have two states in $J^{PC}=1^{+-}$ given by
\begin{eqnarray}
|1^{+-}_a\rangle = \frac{1}{\sqrt{2}} [ |1 1 0\rangle - |1 0 1\rangle ]\ , \quad  |1^{+-}_b\rangle =   |1 1 1\rangle\ ,
\end{eqnarray}
where another labeling, $a$ and $b$, has been introduced to distinguish the two.
Here in $J=1$ channel again, these two states are not the mass eignestates as there is mixing among them through the color-spin
interaction.  Their linear combinations which diagonalize
the hyperfine masses are the mass eigenstates and, therefore, can be identified as the physical states.

\section{Color-spin interactions}
\label{sec:color-spin}

The color-spin interaction provides
a simple way to calculate the {\it mass splittings} among hadrons with the same flavor content, once the
wave functions are constructed in terms of quark fields.  But, in its application to the hadron masses instead of the splittings,
one may need to consider additional terms like color-electric and constant terms which
are often neglected in common practices.
In this section, we want to look at this aspect more carefully
in order to explain why
the color-spin interaction works well for the mass splittings but not for the masses themselves.


\begin{table*}[t]
\centering
\begin{tabular}{c|c|c|c|c}  \hline\hline
      Baryon     & $M_{expt}$ & $M_{H}$ from Eq.(\ref{mass}) &$M_{H}$ from Eq.(\ref{mass add}) &$M_{H}$ from Eq.(\ref{mass gap})\\ \hline
$N$              & 940                 & 844     & 940 (input)  & 940 (input)\\
$\Delta$         & 1232                & 1136    & 1232 (input) & 1232 (input)\\ \hline
$\Sigma$         & 1193                & 1080    & 1182       & 1182\\
$\Lambda$        & 1116                & 1014    & 1116 (input) & 1116 (input)\\
$\Sigma^* $      & 1385                & 1273    & 1375       & 1375\\
$\Xi$            & 1320                & 1223    & 1330       & 1320 (input)\\
$\Xi^*$          & 1531                & 1415    & 1522       & 1513\\ \hline
$\Sigma_c$       & 2453                & 2166    & 2276       & 2438 \\
$\Lambda_c$      & 2286                & 2014    & 2124       & 2286 (input)\\
$\Sigma_c^*$     & 2518                & 2230    & 2340       & 2502\\ \hline \hline
\end{tabular}
\caption{Comparison between baryon masses from PDG and the ones calculated in the three different ways. (See also the text for detail.)
All masses are given in MeV.
The masses in the third column calculated using the color-spin interaction only where
the coupling strength, $v_0^{}$, is fitted from the $\Delta-N$ mass
difference. In obtaining
the fourth column, the additional parameters $v_1$ and $v_2$ in Eq.(\ref{color add})
are fitted by using the $N,\Lambda$ masses
while $v_0$ is still fitted from the $\Delta-N$ mass difference.
The fifth column shows the calculated the masses of $\Sigma$, $\Sigma^* $, $\Xi^*$, $\Sigma_c$, $\Sigma_c^*$, when other
resonance masses are used as inputs in the context of Eq.~(\ref{mass gap}).}
\label{baryon masses}
\end{table*}

The color-spin interaction takes the following simple form~\cite{DeRujula:1975qlm,Keren07,Silve92,GR81}
\begin{equation}
V = \sum_{i < j} v_0^{}\, \lambda_i \cdot \lambda_j \, \frac{J_i\cdot J_j}{m_i^{} m_j^{}}\ .
\label{color spin}
\end{equation}
Here $\lambda_i$ denotes the Gell-Mann matrix for the color SU(3), $J_i$ the spin,
and $m_i^{}$ the constituent mass of the $i$-th quark. The parameter $v_0$ represents
the strength of the color-spin interaction to be fitted from the hadron masses.
It is common practice to write down the hadron masses formally by
\begin{equation}
M_H \sim \sum_{i} m_i^{} + \langle V \rangle\ ,
\label{mass}
\end{equation}
where the hyperfine mass $\langle V \rangle$ is the expectation value of the color-spin interaction
evaluated with respect to an appropriate hadron
wave function of concern.  Phenomenologically, the successful aspect of the color-spin interaction is that
the mass splitting among hadrons (with the same flavor content) can be estimated simply
by the difference in the hyperfine masses~\cite{Kim:2014ywa,Lipkin:1986dx,Lee:2009rt},
\begin{equation}
\Delta M_H \sim \Delta \langle V \rangle\ .
\label{mass gap}
\end{equation}
This relation can be tested, for instance, in the baryon sector.
In this case, one can fix the strength, $v_0$ in Eq.(\ref{color spin}), for example, from the $\Delta -N$
mass difference, and using the standard values for the constituent quark masses, $m_u=330$ MeV, $m_s=500$ MeV, $m_c=1500$ MeV,
one can calculate the mass differences among other hadrons like $\Sigma-\Lambda$, {\it etc}, through Eq.~(\ref{mass gap}).
The agreement with actual mass splittings is quite good, with only $\sim$ 10 MeV error. (See Table VI in Ref.~\cite{Kim:2014ywa}.)

But, if one calculates the actual masses using the formal relation, Eq.(\ref{mass}), the results do not
agree with the experimental masses as one can see from the third column in Table~\ref{baryon masses}.
The disagreement is about 300 MeV for charmed baryons, but for the rest, it is about 100 MeV.
So, the disagreement even has a flavor dependence also. If one sticks to the relation Eq.(\ref{mass}) firmly,
one may attribute this disagreement to the parameters involved, $v_0$ and the quark masses.
One can then adjust these parameters to
reproduce the baryon masses and use them to calculate the masses of other hadrons~\cite{Maiani:2004vq}.
But our standpoint is that there are physical reasons why Eq.(\ref{mass}) fails in reproducing the masses.

Physically, the interaction between two quarks inside a hadron can have two different sources,
one-gluon-exchange potential $V_{OGE}$ and instanton-induced potential $V_{ins}$~\cite{OT89,Oka:1990vx}.
The color-spin interaction is a common ingredient in these potentials but there are additional terms,
the color-electric type, $\sim \lambda_i \cdot \lambda_j$, and the constant term.
Therefore, it may be more natural to think that the disagreement in the masses
above may come from these types of interactions instead of adjustment of the parameters involved in Eq.(\ref{mass}).

Effectively, the additional potentials, being the color-electric type and the constant shift, can be parameterized as
\begin{equation}
V_{add} = \sum_{i < j} v_1^{}\, \frac{\lambda_i \cdot \lambda_j}{m_i^{} m_j^{}} + v_2\ ,
\label{color add}
\end{equation}
with the additional parameters $v_1$ and $v_2$ to be fitted from some of the baryon masses.
Then the baryon mass is given by
\begin{equation}
M_H \sim \sum_{i} m_i^{} + \langle V \rangle + \langle V_{add} \rangle\ .
\label{mass add}
\end{equation}
Now by fitting the additional parameters $v_1$ and $v_2$ to the masses of $N$ and $\Lambda$, for example, we obtain
other masses given in the fourth column of Table~\ref{baryon masses}. The results
are better than the ones determined from the color-spin interaction only, though there are still
disagreement in the charmed baryons more than 160 MeV or so.
One can adjust the parameters at this stage to reproduce the
charmed baryon masses using Eq.~(\ref{mass add}).  In particular, by changing $m_c$ from 1500 MeV to 1670 MeV,
we come out with much better fits to the charmed baryon masses, namely, $M_{\Sigma_c}=2450$ MeV, $M_{\Lambda_c}=2294$ MeV,
$M_{\Sigma^*_c}=2508$ MeV.  Although the results are nice but, at the same time, we are facing the results that
are somewhat sensitive to the input parameters. Then it is questionable
whether this new set of parameters can be consistently applied to other hadrons altogether.

What is interesting is that the additional potentials $\langle V_{add} \rangle$ in Eq.(\ref{mass add}) cancel in the mass splittings among
hadrons with the same flavor content.
Since two quarks inside a baryon are always in the color state ${\bar{\bm{3}}}_c$, the expectation value of the $\lambda_i \cdot \lambda_j$ term is
the same for all the baryons considered in Table~\ref{baryon masses}. The constant term is also the same for all the baryons.
The flavor dependence comes from the quark masses only.
Hence, the relation for the mass splitting, Eq.(\ref{mass gap}),
still holds regardless of the additional potentials as long as one considers hadrons with the same flavor content.

Therefore, a better estimation for the baryon masses can be made if one calculates the mass splitting, $\Delta M_H$, via Eq.~(\ref{mass gap}),
and estimates one of the baryon masses involved in the splitting using other baryon mass as an input.
For example, since $\Sigma -\Lambda$ mass difference is well fitted to
$\Delta \langle V \rangle=\langle V \rangle_{\Sigma}-\langle V \rangle_{\Lambda}$,
one can estimate the $\Sigma$ mass by $M_{\Sigma} = M_{\Lambda}+ \Delta \langle V \rangle$ using the experimental
mass $M_{\Lambda}$ as an input.
Likewise, using $M_N, M_{\Delta},M_{\Xi},M_{\Lambda_c}$ as inputs, we estimate the other baryon masses and they are given
in the fifth column in Table~\ref{baryon masses}, which fit very well with their experimental values.
The reason why this method gives the best estimation among the three calculations in Table~\ref{baryon masses}
is simply because the mass splittings fit to hyperfine
mass splittings quite well without suffering from the additional potentials. To check the sensitivity to the input parameters,
we also change $m_c$ from 1500 MeV to 1670 MeV as above, and obtain $M_{\Sigma_c}=2442$, $M_{\Sigma^*_c}=2500$ MeV. These numbers
are not
so different from the ones in the fifth column in Table~\ref{baryon masses}, indicating that the results are not so sensitive to the input parameters.
One may argue that this way of estimating the baryon masses is not so persuasive
as the half of the resonances are used as inputs. The upshot that we are driving at is that the color-spin interaction can be
powerful only when it is used in the context of the mass splittings.

More importantly, this way of estimating the masses, relying only on hyperfine mass splittings, can be
very powerful in its application to the $X$ resonances. As all the $X$ resonances are assumed to have the same flavor content, once
the mass splittings are calculated via Eq.~(\ref{mass gap}), we can estimate
all the masses using only one resonance mass as an input.  Based on the experience in the baryon sector as well as in the meson sector,
this method relying on the mass splittings can determine the masses quite reliably.

\section{Results and discussion}
\label{sec:results}


\begin{table*}[t]
\centering
\begin{tabular}{c|l}  \hline\hline
States, $ |J^{PC} \rangle $
& ~Hyperfine mass $\langle J^{PC} | V | J^{PC}\rangle_{q_1^{} q_2^{} {\bar q}_3^{} {\bar q}_4^{}}$  \\[1mm]
\hline
$|0^{++}_a \rangle $ &
$\displaystyle 2 v_0^{} \left [ \frac{1}{m_1^{} m_2^{}} + \frac{1}{m_3^{} m_4^{}}
\right ] $ \\[3mm]
$|0^{++}_b \rangle $ &
$\displaystyle -\frac{2}{3} v_0^{} \left [ \frac{1}{m_1^{} m_2^{}} + \frac{1}{m_3^{} m_4^{}}
- \frac{1}{m_1^{} m_3^{}} - \frac{1}{m_1^{} m_4^{}} - \frac{1}{m_2^{} m_3^{}} - \frac{1}{m_2^{} m_4^{}}
\right ] $ \\[3mm]
Mixing [$|0^{++}_a \rangle, |0^{++}_b \rangle $] &
$ \displaystyle \frac{1}{\sqrt{3}} v_0^{} \left [ \frac{1}{m_1^{} m_3^{}} - \frac{1}{m_1^{} m_4^{}}
- \frac{1}{m_2^{} m_3^{}} + \frac{1}{m_2^{} m_4^{}} \right ] $ \\[3mm] \hline
$|1^{++} \rangle $ &
$\displaystyle \frac{1}{3} v_0^{} \left [ \frac{2}{m_1^{} m_2^{}}+\frac{2}{m_3^{} m_4^{}} - \frac{1}{m_1^{} m_3^{}}
+\frac{1}{m_1^{} m_4^{}} + \frac{1}{m_2^{} m_3^{}} - \frac{1}{m_2^{} m_4^{}}
\right ] $  \\[3mm]
$|1^{+-}_a \rangle $ &
$\displaystyle \frac{1}{3} v_0^{} \left [ \frac{2}{m_1^{} m_2^{}} + \frac{2}{m_3^{} m_4^{}} + \frac{1}{m_1^{} m_3^{}}
-\frac{1}{m_1^{} m_4^{}} - \frac{1}{m_2^{} m_3^{}} + \frac{1}{m_2^{} m_4^{}}
\right ] $  \\[3mm]
$|1^{+-}_b \rangle $ &
$\displaystyle -\frac{2}{3} v_0^{} \left [ \frac{1}{m_1^{} m_2^{}}+\frac{1}{m_3^{} m_4^{}}-\frac{1}{2 m_1^{} m_3^{}}
-\frac{1}{2 m_1^{} m_4^{}}-\frac{1}{2 m_2^{} m_3^{}} - \frac{1}{2 m_2^{} m_4^{}}
\right ] $  \\[3mm]
Mixing [$|1^{+-}_a \rangle, |1^{+-}_b \rangle $] &
$\displaystyle -\frac{2}{3} v_0^{} \left [ \frac{1}{m_1^{} m_3^{}} - \frac{1}{m_2^{} m_4^{}}
\right ] $ \\[3mm] \hline
$|2^{++} \rangle $ &
$\displaystyle -\frac{2}{3} v_0^{} \left [ \frac{1}{m_1^{} m_2^{}}+\frac{1}{m_3^{} m_4^{}}
+\frac{1}{2 m_1^{} m_3^{}} + \frac{1}{2 m_1^{} m_4^{}} + \frac{1}{2 m_2^{} m_3^{}} +
\frac{1}{2 m_2^{} m_4^{}} \right ] $ \\[3mm]
\hline\hline
\end{tabular}
\caption{The formulas for the hyperfine masses are provided here, which are obtained from expectation
values of the color-spin interaction with respect to the states indicated in the first column.
The formulas presented here are for a general flavor combination,
$q_1^{} q_2^{} {\bar q}_3^{} {\bar q}_4^{}$.  For the $X$ resonances in this work, one needs to replace simply $m_1=m_c,m_2=m_u,m_3=m_c,m_4=m_u$.}
\label{hyperfine mass formula}
\end{table*}

Now we present our calculation for the hyperfine masses which can be used to estimate the mass splittings among the $X$ resonances.
The hyperfine mass is the expectation value of Eq.~(\ref{color spin}) with respect to the corresponding state
that we have introduced in Sec.~\ref{sec:wf}. Then the color and spin parts need to be calculated separately.
For the color part, since the diquark and antidiquark are in ${\bar{\bm{3}}}_c$, ${\bm 3}_c$ respectively, one can easily evaluate the expectation value of
the terms like $\lambda_1 \cdot \lambda_2$ and $\lambda_3 \cdot \lambda_4$. For the other combinations like $\lambda_1 \cdot \lambda_3$ and
$\lambda_2 \cdot \lambda_3$, etc, one can rearrange the wave functions from the diquark-antidiquark ([12][34]) basis into the ([13][24]) basis
or the ([14][23]) basis and evaluate the expectation values straightforwardly.
One can take the similar steps to calculate the spin parts also. (See Ref.~\cite{Kim:2014ywa} for technical details.)

Table~\ref{hyperfine mass formula} shows our results for the hyperfine masses
evaluated for the states, $|0^{++}_a\rangle$, $|0^{++}_b\rangle$, $|1^{++} \rangle$, $|1^{+-}_a\rangle$, $|1^{+-}_b\rangle$, $|2^{++}\rangle$,
as well as the mixing terms appearing in spin-0, spin-1 channels through $\langle 0^{++}_a|V|0^{++}_b\rangle$, $\langle 1^{+-}_a|V|1^{+-}_b\rangle$.
The origin of the mixing
is that, even though the diquark ([12]) and antidiquark ([34]) are in definite color and spin states, the quark pairs like [13],[14],[23],[24], are not
in definite color and spin states.  The color-spin interaction between two quarks in such a pair gives nonzero contribution in
the matrix elements of $\langle 0^{++}_a|V|0^{++}_b\rangle$, $\langle 1^{+-}_a|V|1^{+-}_b\rangle$.  One can explicitly see this
from our formulas for the mixing terms in Table~\ref{hyperfine mass formula}, where only the terms related to such a pair appear.
Note, the state $|1^{++} \rangle$ can not mix
with $|1^{+-}_a\rangle$ or $|1^{+-}_b\rangle$ because they are in the different states in charge conjugation.
Or one can explicitly check this in our approach by showing that $\langle 1^{++}|V|1^{+-}_a\rangle = \langle 1^{++}|V|1^{+-}_b\rangle=0$.

In our numerical calculations, the input parameters are $m_c$, $m_u$, and $v_0$ in Eq.~(\ref{color spin}). For a consistent treatment
with our earlier work on the $D$ and $B$ meson excited states~\cite{Kim:2014ywa}, we take the same
parameters, namely $m_c=1500$ MeV, $m_u=330$ MeV, $v_0^{} \sim(-193)^3$~MeV$^3$. In Ref.~\cite{Kim:2014ywa}, the strength $v_0$
was fixed by the mass difference $D^{*0}_2(2463)-D^{*0}_0(2318)$ in the four-quark formalism for the $D$ and $B$ meson excited states.
This value is slightly different from the one fitted from the $\Delta-N$ mass difference in the baryon sector.  But $v_0$
becomes larger when it is fitted from the $\rho-\pi$ mass difference.  Thus, the extracted value for $v_0$ has some
dependence on the number of quarks consisting the hadrons and this is a limitation of the color-spin interaction at the moment.
Then the question is which $v_0$ to use in our work.
It may be quite realistic to use the $v_0$ determined from the $D$ meson spectrum in Ref.~\cite{Kim:2014ywa}
because both formalisms are commonly for four-quark systems.
Now, the only step left in getting actual numbers for the hyperfine masses is to identify $m_1=m_c,m_2=m_u,m_3=m_c, m_4=m_u$
in Table~\ref{hyperfine mass formula}.

In the spin-0 channel, the hyperfine masses for the two possible states, $|0^{++}_a\rangle$, $|0^{++}_b\rangle$
are obtained from the corresponding formulas in Table~\ref{hyperfine mass formula}.
As there are the mixing terms between the two states, one has to diagonalize the matrix
to calculate the physical hyperfine masses which then correspond to the physical states.
We denote the physical states with the capital letters in the subscripts as $|0^{++}_A \rangle, |0^{++}_B \rangle$.
\begin{eqnarray}
\begin{array}{c|lr}
& |0^{++}_a\rangle & |0^{++}_b \rangle \\
\hline
|0^{++}_a \rangle & -58.0 & -23.1\\
|0^{++}_b \rangle & -23.1 & -46.0
\end{array}
\quad
&\rightarrow&
\quad
\begin{array}{c|lr}
 & |0^{++}_A \rangle & |0^{++}_B \rangle \\
\hline
|0^{++}_A \rangle & -28.1 & 0.00\\
|0^{++}_B \rangle & 0.00  & -75.9
\nonumber
\end{array}
\end{eqnarray}
Thus, in this spin-0 channel, the hyperfine masses for physical states, $|0^{++}_A \rangle$ and $|0^{++}_B \rangle$, are the followings,
\begin{eqnarray}
\langle 0^{++}_A|V|0^{++}_A\rangle &=& -28.1~{\rm MeV}\ ,\nonumber \\
\langle 0^{++}_B|V|0^{++}_B\rangle &=& -75.9~{\rm MeV}\ .
\label{spin0V}
\end{eqnarray}

Similarly, using corresponding formulas for the spin-1 channel in Table~\ref{hyperfine mass formula},
we can calculate
the hyperfine masses for the two possible configurations, $|1^{+-}_a\rangle$ $|1^{+-}_b\rangle$, as well as their
mixing terms. Then after the diagonalization, we obtain the followings.
\begin{eqnarray}
\begin{array}{c|lr}
& |1^{+-}_a\rangle & |1^{+-}_b \rangle \\
\hline
|1^{+-}_a \rangle & -32.7 & -41.8\\
|1^{+-}_b \rangle & -41.8 & -13.4
\end{array}
\quad
&\rightarrow&
\quad
\begin{array}{c|lr}
 & |1^{+-}_A \rangle & |1^{+-}_B \rangle \\
\hline
|1^{+-}_A \rangle & 19.9 & 0.00\\
|1^{+-}_B \rangle & 0.00  & -65.9
\nonumber
\end{array}
\end{eqnarray}
Again, we have denoted the physical states with the capital letters in the subscripts, $|1^{+-}_A \rangle, |1^{+-}_B \rangle$.
Thus, in this spin-1 channel, the hyperfine masses for physical states are the followings,
\begin{eqnarray}
\langle 1^{+-}_A|V|1^{+-}_A\rangle &=& 19.9~{\rm MeV}\ ,\nonumber \\
\langle 1^{+-}_B|V|1^{+-}_B\rangle &=& -65.9~{\rm MeV}\ .
\label{spin1V}
\end{eqnarray}

There is one more state in spin-1, $|1^{++}\rangle$,
and the spin-2 state, $|2^{++}\rangle$, which do not mix with the other states.
Their hyperfine masses are calculated to be
\begin{eqnarray}
\langle 1^{++}|V|1^{++}\rangle &=& -5.96~{\rm MeV}\ ,\nonumber \\
\langle 2^{++}|V|2^{++}\rangle &=& 52.0~{\rm MeV}\ .
\label{spin12V}
\end{eqnarray}

Now using Eq.(\ref{mass gap}), the mass splittings among theses resonances can be estimated from
the differences in the hyperfine masses given in Eqs.~(\ref{spin0V}),(\ref{spin1V}), (\ref{spin12V}).
For example, the mass difference between the states, $|2^{++} \rangle, |1^{++} \rangle$ is given by
$\langle 2^{++}|V|2^{++}\rangle-\langle 1^{++}|V|1^{++}\rangle=57.96~{\rm MeV},$
As we have emphasized in Sec. III, the mass splittings calculated this way are quite insensitive to the input parameters and do not suffer
from the additional potentials.

Since all these resonances have the same flavor content, we can take the mass of one resonance
as an input in order to estimate the other masses.
The well-known resonance $X(3872)$, which can be identified with the state $|1^{++}\rangle$ because of its quantum numbers,
would be the best candidate for the reference mass. Then, combining with the mass splittings, one can estimate
all the masses of the states, $|0^{++}_A \rangle$, $|0^{++}_B \rangle$, $|1^{+-}_A \rangle$,
$|1^{+-}_B \rangle$,$|1^{++}\rangle$,$|2^{++}\rangle$. For example, the mass of the spin-2 state can be calculated through
\begin{eqnarray}
M(J=2)&=&M_X (3872)+\langle 2^{++}|V|2^{++}\rangle-\langle 1^{++}|V|1^{++}\rangle
\nonumber \\
&=& 3871.7+52-(-5.96)=3929.7~{\rm MeV}\ .
\end{eqnarray}
Similarly we can estimate the masses of the other states.

The states corresponding to the quantum numbers and their calculated masses are summarized as follows,
\begin{eqnarray}
|0^{++}_A \rangle : && \quad 3849.5 ~{\rm MeV}\ ,\nonumber \\
|0^{++}_B \rangle : && \quad 3801.8 ~{\rm MeV}\ ,\nonumber \\
|1^{++} \rangle :   && \quad 3871.7 ~{\rm MeV}\ , ({\rm input})\ ,\nonumber \\
|1^{+-}_A \rangle : && \quad 3897.7 ~{\rm MeV}\ ,\nonumber \\
|1^{+-}_B \rangle : && \quad 3811.7 ~{\rm MeV}\ ,\nonumber \\
|2^{++} \rangle :   && \quad 3929.7 ~{\rm MeV}\ .
\end{eqnarray}
If we compare these masses with the $X$ resonances in Table~\ref{X resonances},
we find that the calculated mass of $|1^{+-}_A \rangle$ is close to $X(3900)^0$ with only 7 MeV difference, and
the mass of $|1^{+-}_B \rangle$, is 11 MeV lower than $X(3823)^0$.
Also the mass of $|2^{++} \rangle$ is found to be only 12 MeV lower than the mass of the resonance $X(3940)$.
Thus, if our framework really works,
we can identify $X(3823)^0$, $X(3900)^0$, $X(3940)$ with the quantum numbers $J^{PC}=1^{+-},1^{+-},2^{++}$.
Of course, future measurements on the quantum numbers can settle the identifying issues raised in this work.

In particular, the quantum number for $X(3823)$ is certainly not settled. It could be $^3D_2$ state of ${\bar c}c$ which
can explain a narrow decay width observed experimentally.
Our four-quark assignment
with $J^{PC}=1^{+-}$ leads to a narrow width also based on observation that its fall-apart decay to $D^* {\bar D}$ is prohibited kinematically.
In addition, there is other approach based on a hadronic molecule~\cite{Patel:2014vua} suggesting that its quantum number $J^{PC}=1^{--}$.
Eventually it will be important to confirm the quantum number of $X(3823)$ by measuring for example
the angular distribution of $X(3823)$ and clarify its nature.

As we have already mentioned, in this calculation, we use the same values for the input parameters as in our previous works~\cite{Kim:2014ywa},
where we develop the four-quark
wave functions for the $D$ and $B$ meson excited states. There, we found some of those resonances fit
very nicely with our picture.  To us, it is quite interesting to see that the masses of the $X$ resonances come out
very close to the experimental values without any parameter tuning at least in the spin-1 and spin-2 channels.
This may provide a consistent treatment on open- and hidden-charm mesons within a four-quark framework.

Currently in PDG, there are no resonances corresponding to the states in the spin-0 channel, $|0^{++}_A \rangle$ and $|0^{++}_B \rangle$.
There could be various reasons for this.  One possibility may be due to their broad widths.
If they decay through a fall-apart mechanism, the decay channel
with low-invariant mass like $D{\bar D}$ can be open and their decay width becomes broad.
If so, these resonances are hard to be measured experimentally.
A similar situation can be seen also in the $B$ meson excited states where there are no resonances observed in the spin-0 channel.
(See Table III in Ref.~\cite{Kim:2014ywa}.)
This was explained in our four-quark model where the spin-0 mesons are expected to have broad widths due
to the presence of the decay modes kinematically favorable. But this scenario does not work for the spin-1 and spin-2 channels
and there are indeed some sharp resonances in PDG in these channels.
It is quite likely that the similar explanation can be applied in the $X$ resonances.
Or alternatively one can look for some other QCD dynamics which might impede the formation of resonances
in the spin-0 channel.
Certainly more works need to be done in future to clarify this issue.

\section{Summary}
\label{sec:summary}

In this work, we have calculated the masses of the $X$ resonances in low-lying region by considering them as tetraquark states.
The color-spin interaction is used to calculate the mass splittings and, using $X(3872)$ as an input, we calculate
the masses corresponding to the quantum numbers $|J^{PC}\rangle = |0^{++}\rangle, |1^{+-}\rangle, |2^{++}\rangle$.
We found that masses of the two states in $|1^{+-}\rangle$ and one state in $|2^{++} \rangle$ fit very nicely
to $X(3823)$, $X(3900)$, $X(3940)$ within $\sim$ 10 MeV.
Future experiments can settle the
issues of identifying the resonances by measuring the quantum numbers of the controversial states.
We stressed that the color-spin interaction is powerful in generating
the mass splittings but not for the mass themselves. To directly calculate the masses of hadrons of concern,
the color-spin interaction may not fully represent the potential among two quarks and one needs the additional
potentials which, however, cancel in the mass splittings.  The input parameters that have been used in the model prediction
were taken to be the same as the ones used in the $D$ and $B$ meson excited states in the
four-quark formalism. Therefore, we believe that our results, if they are confirmed by the measurement of the $X$ quantum numbers,
can provide a consistent picture for the open- and hidden-charm mesons in this four-quark formalism.

\acknowledgments

\newblock
The work of H.Kim was supported by Basic Science Research Program through the National Research Foundation of Korea(NRF)
funded by the Ministry of Education(Grant No. 2015R1D1A1A01059529).
The work of K.S.Kim was supported by the National Research Foundation of Korea
(Grant No. 2015R1A2A2A01004727).
The work of D.Jido was partly supported by Grants-in-Aid for Scientific Research from JSPS (25400254).
The work of M.Oka was partly supported by Grants-in-Aid for Scientific Research from JSPS (25247036).

\end{document}